\documentclass[%
reprint,
onecolumn,
superscriptaddress,
amsmath,amssymb,
aps,
pra,
longbibliography,
]{revtex4-2}

\usepackage{graphicx}
\usepackage{dcolumn}
\usepackage{bm}
\usepackage[caption=false]{subfig}
\usepackage{amssymb}
\usepackage{mathtools}
\newcommand{\volsym}{\rlap{\kern.08em--}V} 
\newcommand\lrp[1]{\left( #1 \right)}

\usepackage{lineno}
\usepackage{float}
\usepackage{listings}

\usepackage{graphicx}
\usepackage{verbatim}   
\usepackage{adjustbox}
\usepackage[english]{cleveref}

\usepackage{color}  
\newcommand{\blue}{\color{black}} 

\newcommand{\black}{\color{black}}

\begin{document}

\preprint{Preprint}

\title{Physics-guided machine learning for wind-farm power prediction: \\Toward interpretability and generalizability}

\author{Navid Zehtabiyan-Rezaie}
\email{zehtabiyan@mpe.au.dk}
\affiliation{Department of Mechanical and Production Engineering, Aarhus University, 8200, Aarhus N, Denmark}
\author{Alexandros Iosifidis}
\email{ai@ece.au.dk}
\affiliation{Department of Electrical and Computer Engineering, Aarhus University, 8200, Aarhus N, Denmark}
\affiliation{Center for Digitalization, Big Data, and Data Analytics, Aarhus University, 8200, Aarhus N, Denmark}
\author{Mahdi Abkar}
\email{abkar@mpe.au.dk}
\affiliation{Department of Mechanical and Production Engineering, Aarhus University, 8200, Aarhus N, Denmark}
\affiliation{Center for Digitalization, Big Data, and Data Analytics, Aarhus University, 8200, Aarhus N, Denmark}


\begin{abstract}
With the increasing amount of available data from simulations and experiments, research for the development of data-driven models for wind-farm power prediction has increased significantly. While the data-driven models can successfully predict the power of a wind farm with similar characteristics as those in the training ensemble, they generally do not have a high degree of flexibility for extrapolation to unseen cases in contrast to the physics-based models. In this paper, we focus on data-driven models with improved interpretability and generalizability levels that can predict the performance of turbines in wind farms. To prepare the datasets, several cases are defined based on the layouts of operational wind farms, and massive computational fluid dynamics simulations are performed. The extreme gradient boosting algorithm is used afterward to build models, which have turbine-level geometric inputs in combination with the efficiency from physics-based models as the features. After training, to analyze the models’ capability in generalization, their predictions for the unseen cases with different operating conditions, inflow turbulence levels, and wind-farm layouts are compared to the Park model and an empirical-analytical Gaussian wake model. Results show that the physics-guided machine-learning models outperform both physics-based models showing a high degree of generalizability, \blue and the machine is not sensitive to the choice of the physics-based guide model\black. 
\end{abstract}

\keywords{Wind-farm modeling; Power prediction; Data-driven modeling; Physics-guided machine learning}
\maketitle


\section{Introduction} \label{Sec:Intro}
Physics-based models are the traditional effective tools to study the performance of wind farms. They lie on a spectrum starting with the low-order analytical wake models (AWMs), ascending to the Reynolds-averaged Navier-Stokes (RANS) models, and ending with the high-fidelity models, e.g., large-eddy simulation (LES). The level of success in modeling the complex and dynamic flow in a wind farm progressively improves by increasing the model's fidelity. LESs in the first place and then RANS simulations can capture many aspects of the physics of flow in wind farms and, therefore, are utilized as computational fluid dynamics (CFD) tools by the wind-energy community \cite{Vermeer2003, Sanderse2011}. However, when it comes to massive simulations of wind farms for real-world applications with different layouts or ambient conditions, the enormous computational costs of CFD tools, as well as their complexity of implementation, pave the way for AWMs to be the preferable tool, especially for the wind-energy industry \cite{Archer2018, PorteAgel2019Review}. The AWMs are derived based on mass and momentum conservation laws with simplifying assumptions, e.g., the shape of the velocity-deficit profile, linear expansion of wake, etc.
They are usually comprised of a single-turbine wake model and are used in conjunction with a superposition technique to take into account the effect of interacting upstream wind turbines  \cite{lissaman1979energy, katic1986simple, voutsinas1990analysis, Niayifar2016}. These models may be applied to any wind farm, existing or new, to give an estimate of its performance. 
However, since they are based on several simplistic physical assumptions, they can fall short in real environments and cause large errors and uncertainties in the optimization algorithms and control strategies in wind-energy projects \cite{Meyers2022}.

Thanks to the availability of data from numerical simulations, wind-tunnel experiments, and the supervisory,
control and data acquisition (SCADA) systems of operational wind farms, and more importantly, based on the high capabilities of data-driven techniques, efforts are made to develop accurate and lightweight models to cover the needs of the wind-energy community. They include machine learning (ML) and reduced-order models (ROMs) for static/dynamic wake modeling (e.g., Refs.  \cite{Iungo2015,Ali2021b,Ti2020}) and power-output prediction (e.g., Refs. \cite{Yin2019, Ti2021}). An overview of the application of different data-driven approaches utilized in studying wind-farm flow and power output can be found in Ref. \cite{zehtab2022}. Without any knowledge of the underlying physics and with a black-box approach, purely data-driven (PDD) models can discover relationships between inputs and outputs. The PDD models' performance degrades when applied to an arbitrary wind farm with a layout, operating conditions of turbines, or inflow characteristics other than those of the farms included in the training and testing phases. To compete with the physics-based models in terms of interpretability and generalizability, attention has shifted toward developing physics-guided data-driven (PGDD) models to predict the performance of wind farms. 

Several recent works have attempted to reveal the ability of PGDD models in predicting wind-farm power output. 
Howland and Dabiri \cite{Howland2019} used a physics-guided statistical wake model to predict the power production of wind farms. Five years of SCADA data from the Summerview onshore wind farm, Canada, were used to train and test the model. The developed model needed two learnable parameters to give the power generation of a waked wind turbine as a weighted summation of the upstream turbines' power. To initialize the wakes, a Gaussian distribution was assumed. The model proved to be more accurate than a physics-based wake model and a standard two-layer artificial neural network (ANN).
Yan \textit{et al.} \cite{Yan2019} trained an ANN model using SCADA data including wind speed, wind direction, and generated power from the Lillgrund offshore wind farm, Sweden. They utilized two geometric inputs of blockage ratio and blocking distance, along with the wind speed, as the features of the ANN model. The idea behind using the geometric features was to characterize the wind-farm layout and wake effect in different wind directions to some extent and, consequently, improve the ML model's generalizability. The blockage ratio for each turbine indicated the fraction of rotor area blocked by the upstream wind turbines, assuming a cylindrical wake, and the blocking distance represented the spacing between the blocked point on the rotor and the upstream blocking turbine. However, the formulation for blocking distance was based on the assumption that the wake behind a turbine recovers at a distance of 20 rotor diameters, which may not completely hold true due to the cumulative wake effect of the turbines. The geometric features were averaged to compute them on the farm level and the output of the model was the power of the farm (not individual wind turbines). They examined the performance of the trained model by applying it to the Nørrekær Enge onshore wind farm, Denmark, with 13 inline turbines, resulting in a mean absolute error (MAE) of 6.4\%. 
Park and Park \cite{Park2019} utilized a combination of a graph representation of the entire wind farm and a graph neural network (GNN) model with an engineering wake model as its basis function to predict the power output of turbines within the wind farm. The physics-guided neural layer utilized a weight-computing function depending on the downstream wake distance and radial wake distance with a fairly close strategy as that in Ref. \cite{Yan2019}. To generate the training data, randomly-generated wind farms were simulated using the flow redirection and induction in steady-state (FLORIS) \cite{floris} utility, and the normalized power outputs of the turbines were computed. The model's performance was analyzed on a uniform layout of 35 turbines with a horizontal and vertical spacing of 5 and 7 rotor diameters, where the model could achieve a farm-level error of 8.49\%, 4.28\%, and 5.99\% for a wind speed of 6 m/s, 9 m/s, and 12 m/s, respectively. They reported farm-level errors of 8.77\%, 6.49\%, and 8.37\% for the same range of wind speed when utilizing a PDD model which was another GNN with a data-induced weight function.
Sun \textit{et al.} \cite{Sun2020} developed an ANN model with consideration of the wake effect based on SCADA data for five selected turbines with almost the same latitude on a hilly terrain in the Shiren onshore wind farm, China. The ML model received the wind speed, yaw angles of turbines, and wake network of the farm corresponding to the wind direction (calculated from the wake model) as its features, and predicted the total power output of turbines. An unseen portion of the dataset of the same farm was used to evaluate the trained model which proved satisfactory performance.
Zhou \textit{et al.} \cite{Zhou2022} compared the performance of several models including the 2D Jensen model \cite{Ge2019} (Model 1), an ANN model (Model 2), and four other ANNs with a PGDD modeling approach (Models 3 -- 6). A small-size SCADA data from 5 turbines in a wind farm located in Jiangsu, China, was utilized to train the models and evaluate their performance after the training phase. Model 2 had several features only from SCADA data including wind speed at the upstream wind turbine, wind direction at the upstream wind turbine, the distance between wind turbines, azimuth angle between turbines, yaw error of the upstream turbine, and the yaw error of the downstream turbine. In Model 3, in addition to the features from SCADA data, a set of features calculated through the physical model was also utilized that included lateral and longitudinal distances of wake, wake influence area, wake influence area proportion, equivalent wind speed, and equivalent wind speed cube. Model 4 had the physical loss included in the ANN's weights optimization process without any features from the physical model. Model 5 was a parallel combination of an ANN and a physical model, where both received SCADA data, and their outputs were aggregated in a weighted network. Model 6 was a data-driven transfer regression with data generated using the low-order physical model. A closer look at the results shows that Model 1 is generalizable as expected, but the average absolute percentage errors of all PGDD models in the prediction phase compared to those of the training phase showed an increase higher than 60\%, even reaching a value of 82\% for Model 6. The PGDD models had a slightly better performance in the prediction phase compared to the 2D Jensen model. \blue In both training and prediction phases, Model 3, guided with physics via its features, outperformed Model 4 which had a physical loss function\black.

A close look at the literature reveals that a key strategy, that is present in the effective and successful procedure followed in the development of the physics-based models, seems to be missing in the data-driven approach. For the physics-based models, the development starts with simple models that are effective and generalizable at the same time, and then, by increasing the complexity level, more accurate models are achieved in such a way that they improve the simpler ones without violating the physical principles. There are gaps in the development of data-driven wind-farm power-prediction models, even in physics-guided ones that can negatively impact the success of data-driven approaches in achieving generalizable models with interpretable results. There is a big contrast between the complexity of the features used in different studies, and the trained models are mostly examined by using an unseen portion of the same training dataset or by applying the models to simple wind farms (e.g., wind farms with few turbines in one row). Therefore, there is doubt about their satisfactory performance when applied to other cases.
Here, we aim at turbine-level power prediction with a PGDD approach and evaluate our trained models by applying them to wind farms with different layouts, turbines' operating conditions, and inflow characteristics. \blue This means that we challenge our ML models from several aspects\black. Such models, if successful, can prove the concept that marching toward generalizable data-driven models for wind-farm power prediction is feasible, and by following a progressive approach, similar to what we already have in the development of the physics-based models, we can achieve generalizable, lightweight, and accurate tools that can extrapolate to other cases, suitable for design, optimization, and real-time applications. \blue Our study advances the state-of-the-art knowledge in data-driven wind-farm power predictions on several fronts in terms of approach and methodology\black. To obtain a sufficient amount of \blue turbine-level \black data for this study, we perform massive CFD simulations on seven different cases with the layouts of operational wind farms. \blue We also investigate the impact of 
introduced physics, by considering two physics-based models with different levels of complexity for the calculation of the turbine-level efficiencies as one of the features of our physics-guided ML models\black. Turbine-level geometric features, as well as the turbines' efficiency from physics-based models, are utilized as the physical features to develop generic data-driven models. Finally, the models' performance is examined on three unseen wind farms with different characteristics compared to the cases present in the training phase. 

The rest of the paper is organized as follows: In Section \ref{Sec:Method}, the methodology of the study is introduced including the data-preparation process and ML models. In Section \ref{Sec:Results}, after studying the impact of the features on the output, ML models' performance on three unseen wind farms is presented and followed by discussions. Finally, the main conclusions from this study are highlighted in Section \ref{Sec:Conclusions}.
\section{Methodology} \label{Sec:Method}
In this section, the data-preparation process is described in detail. Then, the utilized ML algorithm, and the features introduced to the ML models are presented.
\subsection{Data preparation} \label{Sec:Data}
The SCADA data of wind farms are rarely available, and they hold a long list of parameters. Moreover, building a model upon SCADA data can be very challenging, especially in the early stage of the development of interpretable and generalizable models for wind-farm power prediction. Here, we use data generated by numerical simulations to have more control over it and to be able to use a sufficient amount of data that are of simpler nature. 
We make the data publicly available as a step toward collaborative model development, making it possible for the community to compare different models with prior ones.

\begin{figure*}[ht]
\centering
\includegraphics[width=0.9\textwidth]{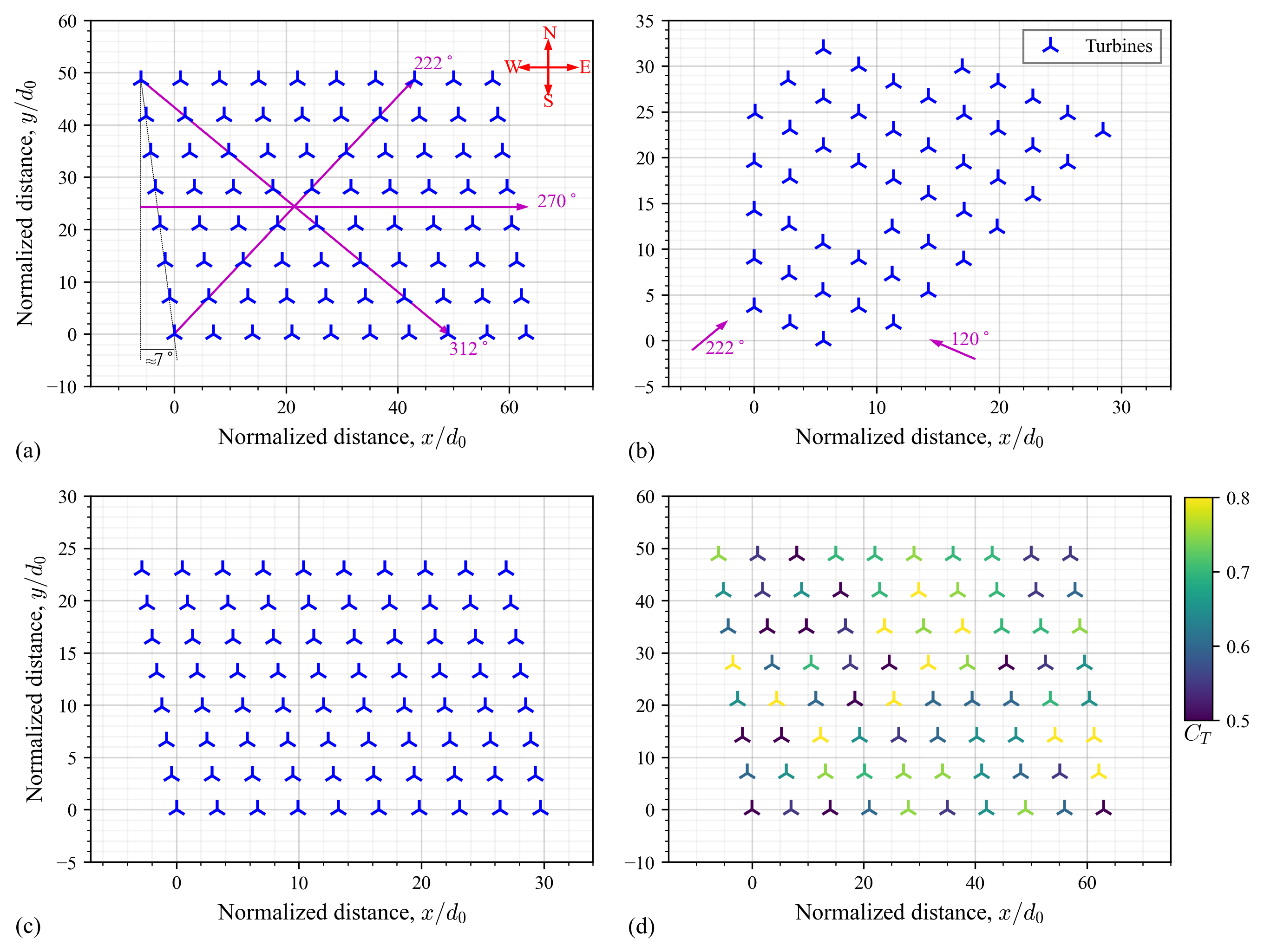}
\caption{Layout of wind farms involved in the data-preparation process (a) HR1 offshore wind farm, (b) Lillgrund offshore wind farm, and (c) Dense HR1. (d) Distribution of thrust coefficient in Case (V). Here, $d_0$ denotes the turbines' rotor diameter that is used to normalize the distances in the corresponding wind farm.}
\label{fig:windFarms}
\end{figure*}

Seven cases based on the layouts of operational wind farms, e.g., Horns rev 1 (HR1) offshore wind farm, Denmark (Figure~\ref{fig:windFarms}(a)), and Lillgrund offshore wind farm, Sweden (Figure~\ref{fig:windFarms}(b)), are defined to generate the required datasets for the data-driven part of our study. The datasets include efficiency values for all turbines in each case for different incoming wind directions with a step of $2^\circ$. \blue The turbine efficiency is the ratio of its power to the extracted power if it was standing upstream\black. Details on the specifications of turbines, farms, and the characteristics of the inflow in each case are provided in Table~\ref{tab:Infor}, where $N_\text{t}$, $d_0$, $z_\text{h}$, $C_T$, $\overline{u}_\text{0}$, and $I_\text{0}$ denote the number of turbines, their rotor diameter and hub height, turbines' thrust coefficient, inlet hub-height velocity, and inlet hub-height turbulence intensity, respectively. 
The wind farms in Cases (I) and (II) have the same layout as the HR1 wind farm. The HR1 wind farm consists of eighty Vestas V-80 2MW turbines with a rhomboid-shaped layout occupying an area of 20 $\text{km}^2$ with a spacing of $7d_0, 9.3d_0$, and $10.4d_0$ between the consecutive turbines for $\theta_\text{wind}=270^\circ$, $222^\circ$, and $312^\circ$, respectively \cite{Wu2015}. In Case (I), a uniform distribution is assumed for the turbines' thrust coefficient with a value of 0.8, while a uniform thrust coefficient of 0.5 is assumed for the turbines in Case (II). To include cases with more severe wake effects due to a shorter inter-turbine spacing, we have created a wind-farm layout which is a dense version of the real HR1 wind farm, as shown in Figure~\ref{fig:windFarms}(c). The dense HR1 has the same number of turbines as the real HR1, but with a spacing of $3.3d_0, 4.4d_0$, and $4.9d_0$ between the consecutive turbines for $\theta_\text{wind}=270^\circ$, $222^\circ$, and $312^\circ$, respectively. Similar to previous cases, two constant values for the turbines' thrust coefficient are considered in the data-preparation process using this layout labeled as Cases (III) and (IV). 
We simulate Case (V) with the same layout as the HR1 wind farm, where we have assigned random thrust coefficients $\in [0.5,0.8]$ to its turbines. In Figure~\ref{fig:windFarms}(d), a colored representation is utilized to illustrate the non-uniform distribution of the thrust coefficient in this case. 
In Case (VI), we simulate the Lillgrund offshore wind farm, assuming a uniform distribution of the thrust coefficient equal to 0.86 based on its turbines' thrust curve \cite{goccmen2016estimation}. The Lillgrund wind farm holds forty-eight Siemens SWT-2.3-93 2.3MW wind turbines. The spacing of consecutive turbines are $3.3d_0, 4.3d_0, 4.8d_0$, $5.9d_0$, and $8.6d_0$ for $\theta_\text{wind}=120^\circ$, $222^\circ$, $180^\circ$, $255^\circ$, and $270^\circ$, respectively. Due to the characteristic gap in its layout, the spacing between some of the turbines can reach $9.1d_0$ for $\theta_\text{wind}=222^\circ$ \cite{LillgrundTech}. The above-mentioned cases are simulated with an inflow turbulence level of $I_\text{0}=7.7\%$. We also simulate Case (VII) which is similar to Case (VI) but with a different hub-height turbulence intensity ($I_\text{0}=15\%$) at the inlet. 

\begin{table*}[ht]
\centering
\caption{Characteristics of different cases in the data-preparation process.}
\label{tab:Infor}
\begin{ruledtabular}
\begin{tabular}{lcccccccccccc}
Case	       &	Layout	             &	$N_\text{t}$   &	$d_0$ 	&	$z_\text{h}$   &	$C_T$   &	$\overline{u}_\text{0}$	  & $I_\text{0}$   & Utilization stage     \\  \hline
Case (I)	   &	HR1 wind farm	     &	80	           &	80 m	&	70 m	       &	0.8	                      &	8 m/s	    & 7.7\%        &	Training \& testing	\\	
Case (II)	   &	HR1 wind farm	     &	80	           &	80 m	&	70 m	       &	0.5	                      &	8 m/s	    & 7.7\%        &	Training \& testing	\\	
Case (III)     &	Dense HR1	         &	80	           &	80 m	&	70 m	       &	0.8	                      &	8 m/s	    & 7.7\%        &	Training \& testing	\\	
Case (IV)	   &	Dense HR1	         &	80	           &	80 m	&	70 m	       &	0.5	                      &	8 m/s	    & 7.7\%        &	Training \& testing	\\
Case (V) 	   &	HR1 wind farm	     &	80	           &	80 m	&	70 m	       &	RND $\in [0.5,0.8]$  	  &	8 m/s	    & 7.7\%        &	Prediction	        \\
Case (VI)	   &	Lillgrund wind farm	 &	48	           &	93 m	&	65 m	       &	0.86	                  &	8 m/s	    & 7.7\%        &	Prediction	        \\	
Case (VII)	   &	Lillgrund wind farm	 &	48	           &	93 m	&	65 m	       &	0.86	                  &	8 m/s	    & 15\%         &	Prediction	       \\	
\end{tabular}
\end{ruledtabular}
\end{table*}

\blue Since we aim to train robust ML models and challenge them in the prediction phase from different aspects, i.e., interpolation of different operating conditions of turbines, as well as extrapolation to unseen operating conditions, wind-farm layout, and inflow characteristics, the prepared datasets segmentation is done with special care and attention.  
Table~\ref{tab:Infor} presents the utilization stages of different cases. Cases (I -- IV) are utilized in the training and testing phases as they construct the parametric space we need by holding 28,800 turbine-level data points. The remaining cases, i.e., Cases (V -- VII), are kept unseen in the training and testing phases to challenge the ML models in the prediction phase from the aspects described earlier\black.

\subsubsection{Wind-farm simulation framework and numerical setup}
Due to the size of the wind farms included in the data-preparation process of our study, LESs under different turbine characteristics and \blue inflow conditions for a \black complete set of incoming wind directions with a step of $2^\circ$ will result in enormous computational costs.  We intend to simulate the full wind farms with all turbines to avoid using any empirical superposition technique in order to capture the wake effects with more accuracy. To control the computational costs of the data-preparation stage and achieve the sufficient amount of data required for effective ML-model training, we use precise numerical simulations through a validated RANS framework.

The Reynolds-averaged governing equations of a turbulent incompressible flow, comprised of the continuity and momentum equations, can be expressed as
\begin{equation} \label{eq:Mass}
\partial_{i} \overline{u}_{i} = 0,
\end{equation}
\begin{equation} \label{eq:Momentum}
\partial_{t} \overline{u}_{i} +\overline{u}_{j} \partial_{j} \overline{u}_{i} = -\frac{1}{\rho } {\partial_{i} \overline{p}} + {\partial_{j}} \left( 2 \nu \overline{S}_{ij} - \overline{R}_{ij}\right)  - \frac{\overline{f}_{i}}{\rho}, \end{equation}
\noindent  where $\overline{u}_{i}$ is the mean velocity, $\overline{p}$ is the mean pressure, $\overline{S}_{ij}$ is the mean rate of strain, $\overline{R}_{ij}$ is the Reynolds stress, and  $\rho$ and $\nu$ are the density and kinematic viscosity of the fluid, respectively. The turbine-induced force  $(\overline{f}_{i})$ is calculated based on the actuator-disk model without rotation, using the velocity at the rotor and the disk-based thrust coefficient $C^\prime_T =C_T /(1-a)^2$, in which, $a$ is the induction factor \cite{calaf2010large}. To formulate the Reynolds stress term, the Boussinesq's hypothesis is utilized as $\overline{R}_{ij}=-2 \nu_\text{T} \overline{S}_{ij} + \frac{2}{3} k \delta_{ij}$, where, $\nu_\text{T}$ is the eddy viscosity, $k$ is the turbulent kinetic energy (TKE), and $\delta_{ij}$ is the Kronecker delta \cite{durbin2018some}. Transport equations are solved for the TKE and its dissipation rate ($\varepsilon$) to compute the eddy viscosity, and similar to our previous works \cite{Eidi2021,Eidi2022} and because of \blue the relatively good agreement with the LES data in predicting the power output\black, we utilize the realizable $k-\varepsilon$ model \cite{shih1995new} for this purpose.

\begin{figure*}[ht]
	\centering
	\includegraphics[width=0.9\textwidth]{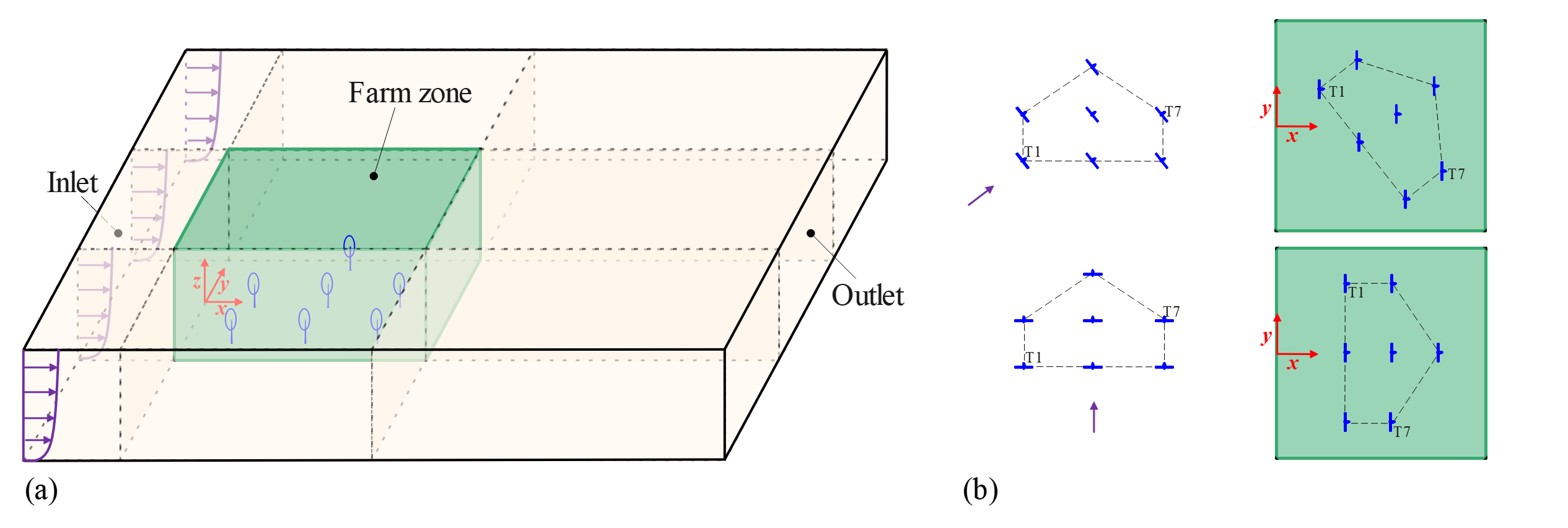}
	\caption{(a) A schematic view of the blocks of the computational domain. The farm zone is highlighted in green (not to scale). (b) A simple example of rotation of the wind turbines inside the farm zone for different wind directions.}
	\label{fig:windFarmBlocks}
\end{figure*}

We adopt the \texttt{simpleFoam} solver, available in OpenFOAM v2012 \cite{openfoamv2112}, to solve the above-mentioned governing equations and model the wake interaction of multiple turbines in the cases introduced earlier under the neutral atmospheric boundary layer (ABL) conditions.
As schematically depicted in Figure~\ref{fig:windFarmBlocks}(a), we create a computational domain with a height of 500 m, comprised of 9 blocks, one of which is the farm zone. The reason behind the consideration of the 8 surrounding blocks is to minimize the effect of the ending boundaries on the flow inside the farm zone. As shown in a very simple example in Figure~\ref{fig:windFarmBlocks}(b), to model the wind farms in different incoming wind directions, the locations of the turbines are updated accordingly, while the inlet boundary is fixed and the turbines are perfectly aligned with respect to the wind direction. We apply the slip boundary conditions at the side boundaries. At the outlet, the zero gradient boundary condition is applied to all quantities and a pressure outlet for the pressure. At the inlet and top boundaries, zero gradient condition is used for the pressure, and for the velocity and turbulence quantities, values corresponding to a logarithmic boundary-layer flow \cite{plate1971aerodynamic} are utilized. 
The computational domain is discretized with 31,217,742, 8,928,804, and 8,280,720 cells for the simulation of cases with the layout of the HR1 wind farm, Dense HR1, and Lillgrund wind farm, respectively. The generated meshes cover the rotor diameter of each turbine by at least 8 points in the spanwise and vertical directions, and a grid-independence test reveals that this resolution can successfully capture the most important characteristics of the wind-turbine wake. While the farm zone is divided uniformly in the $x$ and $y$ directions, stretching grids are utilized in the surrounding blocks to reduce computational costs. 

\begin{figure*}[ht]
\includegraphics[width=0.87\textwidth]{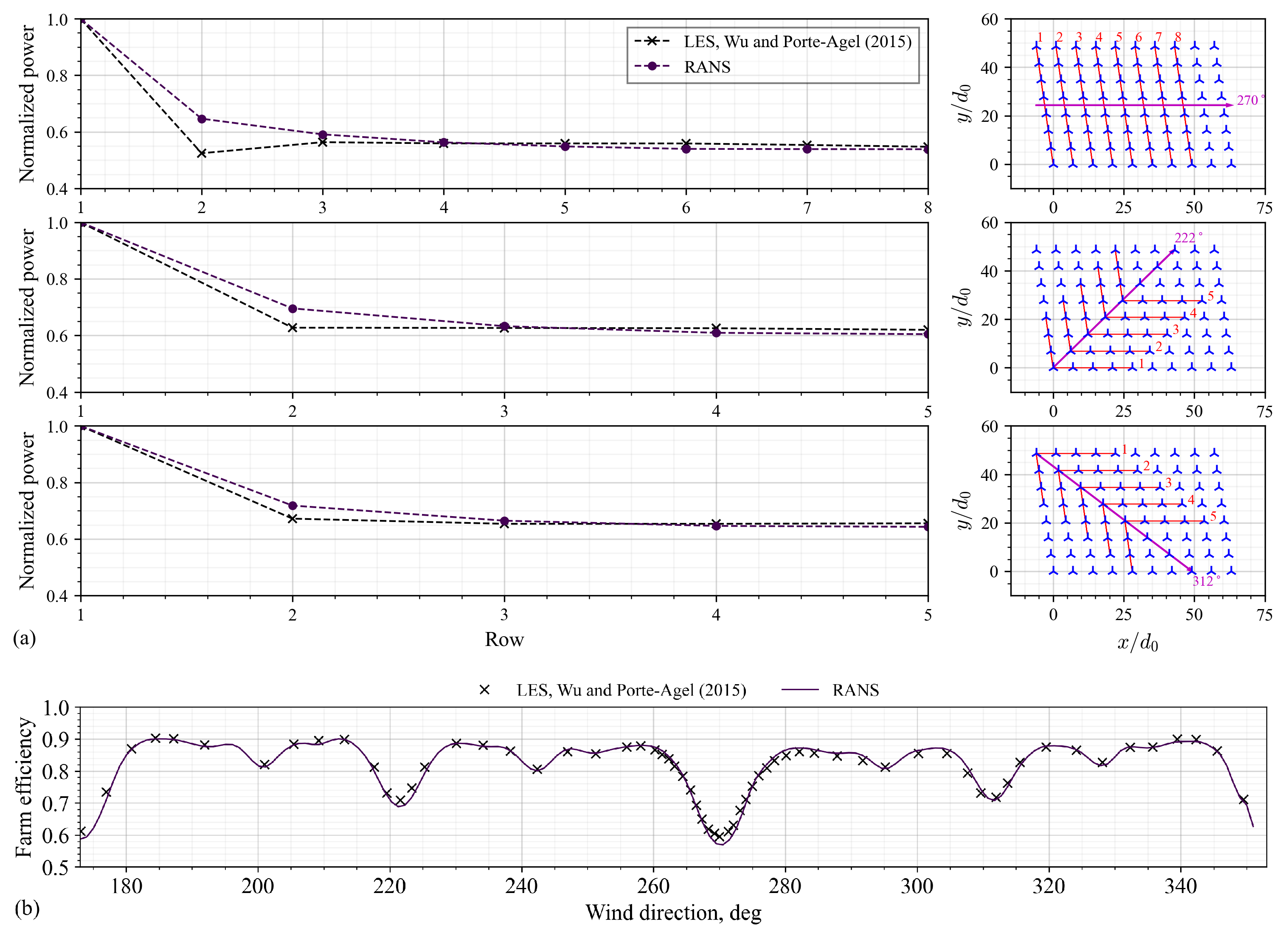} 
\caption{RANS framework's predictions in the simulation of the HR1 wind farm for a hub-height velocity and turbulence intensity of 8 m/s and 7.7\%.  (a) Left: The normalized mean power of turbines' rows at three different wind directions compared to LES data \cite{Wu2015}. Right: The turbines marked with the red line are used to calculate the average power of each row for the three wind directions, and the average value of the first row is used as the reference in the calculation of the normalized power. (b) The efficiency of the HR1 wind farm as a function of incoming wind direction compared to LES results \cite{Wu2015}.} 
\label{fig:ransVal}

\end{figure*}

To validate the RANS framework, in Figure~\ref{fig:ransVal}(a), we compare the normalized power of several selected rows for 3 incoming wind directions with LES results from the study of Wu and Port\'e-Agel \cite{Wu2015}. In Figure~\ref{fig:ransVal}(b), the normalized power of the HR1 wind farm is compared to LES results, for a complete set of incoming wind directions. The comparisons show a relatively good agreement with the LES data, therefore, the RANS framework is utilized to simulate the defined cases to establish the datasets. 

\subsection{Machine-learning model}
\begin{figure*}[ht]
	\centering
	\includegraphics[width=0.75\textwidth]{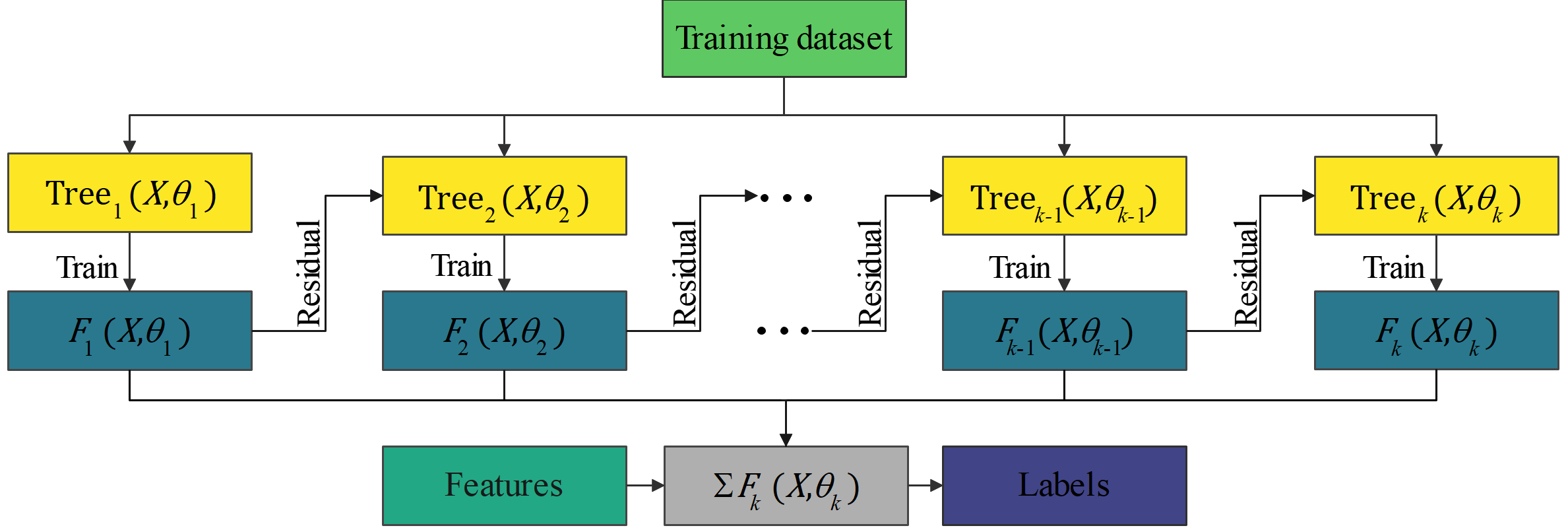}
	\caption{The boosting-learning algorithm.}
	\label{fig:xgBoost}
\end{figure*}

Due to its high speed and effectiveness, we use the extreme gradient boosting (XGBoost) algorithm to build our ML models. The XGBoost model is a form of ensemble learning in which multiple learners are combined to achieve a strong model in the final. By following this procedure, one can start with simple weak learners, and improve the model step by step. As shown in Figure~\ref{fig:xgBoost}, the XGBoost model is based on the gradient-boosting strategy \cite{chen2016xgboost}, in which the initial learner is a weak one, and by checking the residuals which is an indication of learning error, the weights are updated to get a strong learner in the final stage.  The XGBoost model has several hyper-parameters that must be chosen by the user. The most important hyper-parameters also called structure hyper-parameters, are 1) \texttt{max\char`_depth} which indicates the maximum depth of the tree model as the XGBoost is a model established with tree model structures, 2) \texttt{min\char`_child\char`_weight} that gives the minimum number of leaf node samples for the branch split, and 3) \texttt{n\char`_estimators}  which is the number of iterations of the tree model. Larger values for these parameters can lead to the high accuracy of training, however, the XGBoost model has regularization terms in its objective function that can be used if the model tends to be over-fitting. 

\subsubsection{Features of machine-learning model}
Our ML models get the features as the inputs and predict the labels which are the efficiency of the turbines. In this study, three different \blue ML models are trained and evaluated. The turbine-level features of the ML models are given in Table~\ref{tab:featuresML} which are comprised of two geometric features \cite{Ghaisas2016} and efficiencies calculated using the Park model and the analytical-empirical model proposed by Niayifar and Port\'e-Agel (NP model) \cite{Niayifar2016}\black. \blue The Park model is extensively utilized in the wind-energy community and is the backbone of several industry-based software, e.g., wind atlas analysis and application program (WAsP). The NP model \cite{Niayifar2016}, implemented in different utilities such as FLORIS, is an extended version of the Gaussian wake model \cite{Bastankhah2014} that has a higher level of included physics compared to the Park model, and is widely utilized in academia. Including two different physics-based models will create the opportunity to compare their accuracy and more importantly, investigate the impact of the choice of physics-based guide model on the performance of our physics-guided ML models\black. 
\begin{table*}[ht]
\centering
\caption{\blue Features of the ML models. Here, the subscript $j$ denotes that the features are at turbine level\black.}
\label{tab:featuresML}
\begin{ruledtabular}
\begin{tabular}{>{\blue}l>{\blue}c>{\blue}c>{\blue}c>{\blue}c>{\blue}c>{\blue}c}
Model	& Abbreviation       &	$BR_j$	             &	$IBD_j$   	&	$\eta_{\text{Park},j}$   &	$\eta_{\text{NP},j}$    \\  \hline
ML model with geometric features only & GM-ML	   &	$\checkmark$	     &	$\checkmark$	           &	-	    &	-\\	
ML model guided with the Park model  & Park-ML	   &	$\checkmark$	     &	$\checkmark$	           &$\checkmark$	&	-\\	
ML model guided with the NP model \cite{Niayifar2016} & NP-ML     &	    $\checkmark$	      &	$\checkmark$           &		-    &	$\checkmark$	  	\\	
\end{tabular}
\end{ruledtabular}
\end{table*}
\subparagraph{Geometric features:} 
We use the geometric features of blockage ratio (BR), and also the inverse blocking distance (IBD) inspired by the original formulation proposed by Ghaisas \textit{et al.} \cite{Ghaisas2016}, to describe how turbine $j$ is spaced with respect to the upstream turbines in different wind directions. 
The blockage ratio for turbine $j$ is defined as

\begin{align}
	BR_j = \frac{1}{A} \int _{(y,z) \in A} \chi dy\, dz, 
	\label{eq:BlockingRatio}
\end{align}
where $\chi = \chi(y,z)$ is a function defined for all nodes on the rotor disc. The function equals 1 if a node at $(y,z)$ is blocked by any upstream turbine, and 0 otherwise. The wakes behind all upstream turbines are assumed to have a cylindrical shape in this formulation. Here, $A$ is the rotor-disc area ensuring $BR_j \in [0,1]$.

Another geometric feature, named inverse blocking distance, that attempts to characterize the spacing between the turbine and the upstream blocking one is originally  formulated as

\begin{align}
	IBD_{\text{org},j} = \frac{1}{A}  \int _{(x,y) \in A} \frac{\chi}{L} dx\, dy,  
        \label{eq:invBlockingDistanceOrg}  
\end{align}
where $L = L(y,z)$ is the distance of each point on the rotor of turbine $j$ relative to the upstream blocking turbine. This formula works for a case in which there is no overlap among the blocked areas of turbine $j$, but in another case where a point on this turbine is blocked by two or more upstream turbines, the equation only accepts one value of $L$. Instead, we use a cumulative form for the inverse blocking distance for turbine $j$ as

\begin{widetext}
\begin{equation}
        IBD_j =  \sum_{k} \frac{1}{A}  \int _{(y,z) \in A} \frac{\chi_{kj}}{L_{kj}} dy\, dz, \forall	k =1, ..., N_\text{t}  \text{  and  }  x_k \leqslant x_j, 
        \label{eq:invBlockingDistance}  
\end{equation}
\end{widetext}
where $L_{kj} = L_{kj}(y,z)$ is the distance of each point on the rotor of turbine $j$ relative to the upstream turbine $k$.  

\begin{figure*}[ht]
	\centering
	\includegraphics[width=0.88\textwidth]{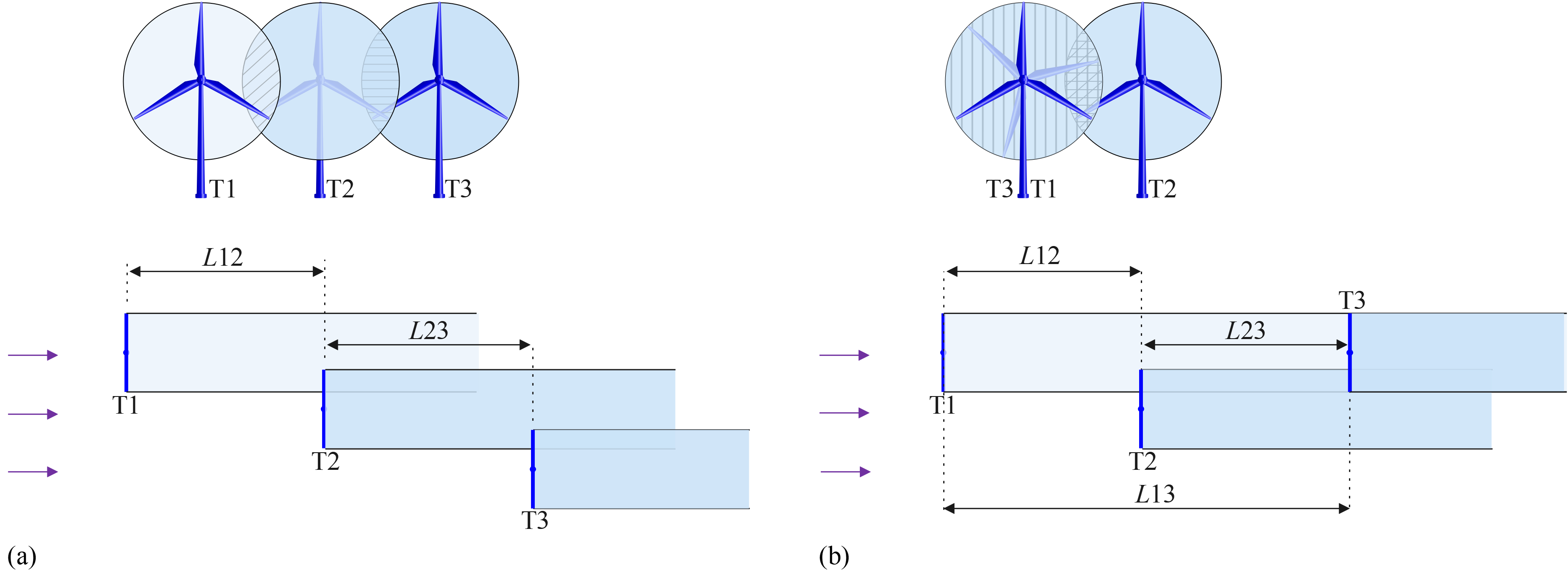}
	\caption{Simple examples of turbine-level calculation of the blockage ratio and inverse blocking distance assuming cylindrical wake behind upstream turbines.}
	\label{fig:BRIBD}
\end{figure*}

To visualize the two geometric features, simple examples of three wind turbines can be seen in Figure~\ref{fig:BRIBD}. In the case in part (a), turbine T3 is partially blocked by turbine T2, e.g., blockage of 10\% of its swept area. Therefore, $BR_3 = 0.1$, and only the effect of T2 is included in the calculation of inverse blocking distance for T3 as $IBD_3 = 0.1 / L_{23}$. In the case in part (b), T3 is affected by both upstream turbines, a partial coverage by T2 (e.g., 10\% of its swept area) and full coverage by T1, therefore, $BR_3 = 1.0$, and the cumulative effect of both T1 and T2 must be included as $IBD_3 = 1.0 / L_{13} + 0.1 / L_{23}$.

We do not expect that the turbine-level geometric features can serve as the only features of an ML model since they do not possess any information on the operating conditions of the turbines or inflow conditions. For an ML model only fed with the geometric features, there will be no difference between the turbines in Cases (I) and (II), or Cases (III) and (IV). Moreover, the assumption of a cylindrical wake behind the upstream turbines is another shortcoming in using geometric features only. \blue However, we train and evaluate the GM-ML model to properly evaluate the geometric features being the only inputs of an ML model. For the other two ML models\black, we include a feature to guide the machine by providing extra physical information that corresponds to different operating conditions of turbines, inflow characteristics, and also wake expansion. As mentioned earlier, \blue two physics-based models with different levels of complexity are \black utilized to generate another input to be combined with the geometric features.

\subparagraph{Turbines' efficiency from Park model:}
The third input included in the features of the Park-ML is the turbine-level efficiencies calculated through the Park model that is based on the Jensen analytical wake model \cite{jensen1983a} with the superposition technique suggested by Katic \textit{et al.} \cite{katic1986simple}. 
In the Park model, a top-hat profile is assumed for the velocity deficit in the wake of a turbine as

\begin{align}
    \frac{\Delta \overline{u}}{ \overline{u}_0 } &= \left( 1 - \sqrt{1 - C_{T}} \right) \bigg/ \left(1 + \frac{2 k_\text{w} x} {d_0} \right)^2, \label{eq:jensen}
\end{align}

\noindent where $k_\text{w}$ is the wake expansion rate and $x$ is the streamwise position. While some studies recommend constant values of 0.04 and 0.05 for $k_\text{w}$ in offshore wind farms \cite{barthelmie2006comparison,barthelmie2007modelling}, the wake expansion rate can be related to the ambient turbulence intensity through $k_\text{w} \approx 0.4I_0$ \cite{pena2016application,abkar2018theoretical}. An energy-deficit superposition is implemented to model the interaction of multiple turbines \cite{katic1986simple} as

\begin{align}
    \overline{u}_j = \overline{u}_0 - \sqrt{\displaystyle\sum_{k} (\overline{u}_0 - \overline{u}_{kj})^2},
    \label{eq:sup1}
\end{align}
\noindent where, $\overline{u}_{kj}$ is the wake velocity of the turbine $k$ at turbine $j$, if the turbine $k$ has a wake action on turbine $j$.

\subparagraph{Turbines' efficiency from NP models:}
\blue The NP-ML model receives the turbine-level efficiencies calculated through the NP model \cite{Niayifar2016} along with the two geometric features. The NP model \cite{Niayifar2016} attempts to extend the Gaussian wake model of Bastankhah and Port\'e-Agel \cite{Bastankhah2014} that was derived based on the conservation of mass and momentum for a wind turbine. The velocity deficit in the turbine wake is given by

\begin{align}
    \frac{\Delta \overline{u}}{ \overline{u}_0 } &= \lrp{ 1 - \sqrt{1 - \frac{C_{T}}{8(k^* x /d_0 + \varepsilon)^2} }} \exp \left\{- \frac{1}{2(k^* x /d_0 + \varepsilon)^2} \lrp{\left(\frac{z-z_\text{h}}{d_0}\right)^2+\left(\frac{y}{d_0}\right)^2 } \right\}, \label{eq:calcDeltaU}
\end{align}
\noindent where, $x$, $y$, and $z$ are coordinates in the streamwise, spanwise, and vertical directions. 
The wake growth rate is assumed to follow a relationship of $k^*= 0.35 I_0 $ \cite{Carbajo2018}. For each turbine, the added turbulence intensity solely from the nearest upstream turbine with the most significant wake impact is included. The wake turbulence intensity is calculated through $I_{\text{w}} = \sqrt{I_0^2+I_+^2}$, where $I_{+} = 0.73 a^{0.8325} I_0^{-0.0325} \left({x}/d_0\right)^{-0.32}$ is the added turbulence intensity \cite{Crespo1996}. To calculate the velocity deficit at the turbine $j$, a superposition technique is utilized as

\begin{align}
    \overline{u}_j = \overline{u}_0 - \displaystyle\sum_{k} (\overline{u}_k - \overline{u}_{kj}), 
    \label{eq:sup2}
\end{align}
\noindent where, $\overline{u}_{k}$ is the inflow velocity at the turbine $k$. For more information, see, e.g., Ref. \cite{Niayifar2016}.
\black
\section{Results and discussion} \label{Sec:Results}
In this section, first, the training phase of the ML models is analyzed, \blue followed by a thorough discussion on the importance of the ML models' features and interpretability of  their results\black. Finally, predictions of the trained model on the unseen cases are evaluated. 

\subsection{Training and hyper-parameters tuning}
As discussed earlier, we divide the datasets of seven cases into two parts and utilize the first part, Cases (I -- IV), in the training and testing phases of our ML models, keeping the second part, Cases (V -- VII), for the next step that is performed to evaluate the ML models' performance when applied to unseen cases, as a measure of their generalizability. 
To train the ML models, the data are scaled before the training process which generally speeds up learning and leads to faster convergence. The datasets for training and testing are randomly selected, holding 80\% and 20\% of the initial dataset, respectively. By using a random search \cite{bergstra2012random} algorithm, the hyper-parameters of the XGBoost models are optimized, as given in Table~\ref{tab:hypers}.

\blue 
The learning curves of the ML models, showing root mean square error versus the number of iterators for both training and testing losses, are analyzed to monitor the convergence. For both physics-guided ML models, the learning curves show a decreasing behavior to a point of stability with a minimal gap between the two final loss values which is a sign of a satisfactory fit. While the learning curve of the GM-ML model has a similar trend, the converged loss is higher than that of the physics-guided ML models. The coefficient of determination $(R^2)$ of the physics-guided ML models in the training and testing phases are 98\% and 97\%, respectively, while these indexes degrade to 93\% and 91\% for the GM-ML model. Proper evaluation of the success level of our ML models in the training phase is crucial for evaluating the models' predictions when applied to unseen cases. To this end, we compare our ML models by using the performance metric of the mean of turbine-level absolute percentage error (\textit{MAPE}) when applied to the training and testing data. The \textit{MAPE} is given by

\begin{align}
       MAPE &= \frac{1}{N_\text{t}} \sum_{j=1}^{N_\text{t}} {\left\lvert \frac{y_{\text{pred},j}-y_{\text{true},j}}{y_{\text{true},j}} \right\rvert} \times 100.
\end{align}

The (\textit{MAPE}$_\text{train}$, \textit{MAPE}$_\text{test}$) averaged for all wind directions are equal to (2.60\%, 2.87\%), (2.27\%, 2.77\%), and (5.76\%, 6.21\%) for the Park-ML, NP-ML, and GM-ML models. Looking at the \textit{MAPE} of the GM-ML model in the training and testing, we can highlight once again the two geometric features alone are not capable enough to result in robust ML-model training as they do not possess information on the physics of the flow and operating conditions. 
\black

\begin{table*}[ht]
\centering
\caption{\blue Tuned hyper-parameter of the XGBoost regression models.\black}
\label{tab:hypers}
\begin{ruledtabular}
\begin{tabular}{l c c >{\blue} c}
Hyper-parameter                   &   GM-ML    &    Park-ML &   NP-ML \\ \hline
Number of estimators              &  800       &  800       &   1600         \\
Maximum tree depth                &  5         &  5         &   6        \\
Sub-sample ratio                  &  0.7       &  0.5       &   0.6    \\
Minimum child weight              &  7         &  5         &   7          \\ 
Fraction of sub-sampled columns   &  0.1       &  1         &   0.9       \\
Learning rate                     &  0.1       &  0.1       &   0.08         \\ 
\end{tabular}
\end{ruledtabular}
\end{table*}

\subsection{Interpreting machine-learning models' predictions}
Interpretability of ML models developed for wind-farm power prediction is becoming an increasingly important requirement. Here, we apply the Shapley additive explanations (SHAP) method proposed by Lundberg and Lee \cite{ lundberg2017SHAP} based on classical Shapley values from game theory  to the entire population of the training dataset to achieve global interpretations on the predictions of the ML models. SHAP is one of the most powerful methods for explaining how an ML model makes individual predictions. However, we need to be cautious while interpreting such results since SHAP only tells what the model is doing within the context of the data on which it has been trained, and it does not necessarily reveal the true relationship between variables and outcomes in the real world. In the SHAP method, an ML model’s predictions can be formulated as a fixed base value plus the summation of features’ SHAP values. For regression models, the base value is the average of the target variable. \blue According to the local accuracy property, a simpler explanation model $g$ is used as the interpretable explanation of the original model $f$ for a specific instance $x$ through the explanation attribution values $\phi$ for each feature $l$ as

\begin{align}
   f(x) = g(x')=\phi_0 + \displaystyle\sum_{l} \phi_l x'_l,
    \label{eq:prop1}
\end{align}
\noindent where the original input $x$ is related to the simplified input through mapping as $x=h_x(x')$, and $\phi_0$ represents the constant value when all simplified inputs are toggled off. SHAP values have other properties including missingness and consistency. The missingness property says that a missing feature gets an attribution of zero, and the consistency property states that if a model changes so that the impact of a feature increases or remains the same, its SHAP value will never decrease. For a more detailed description of the SHAP method, see, e.g., Refs. \cite{lundberg2017SHAP,lundberg2020local2global}.
\black

\begin{figure*}[ht]
	\centering
	\includegraphics[width=1\textwidth]{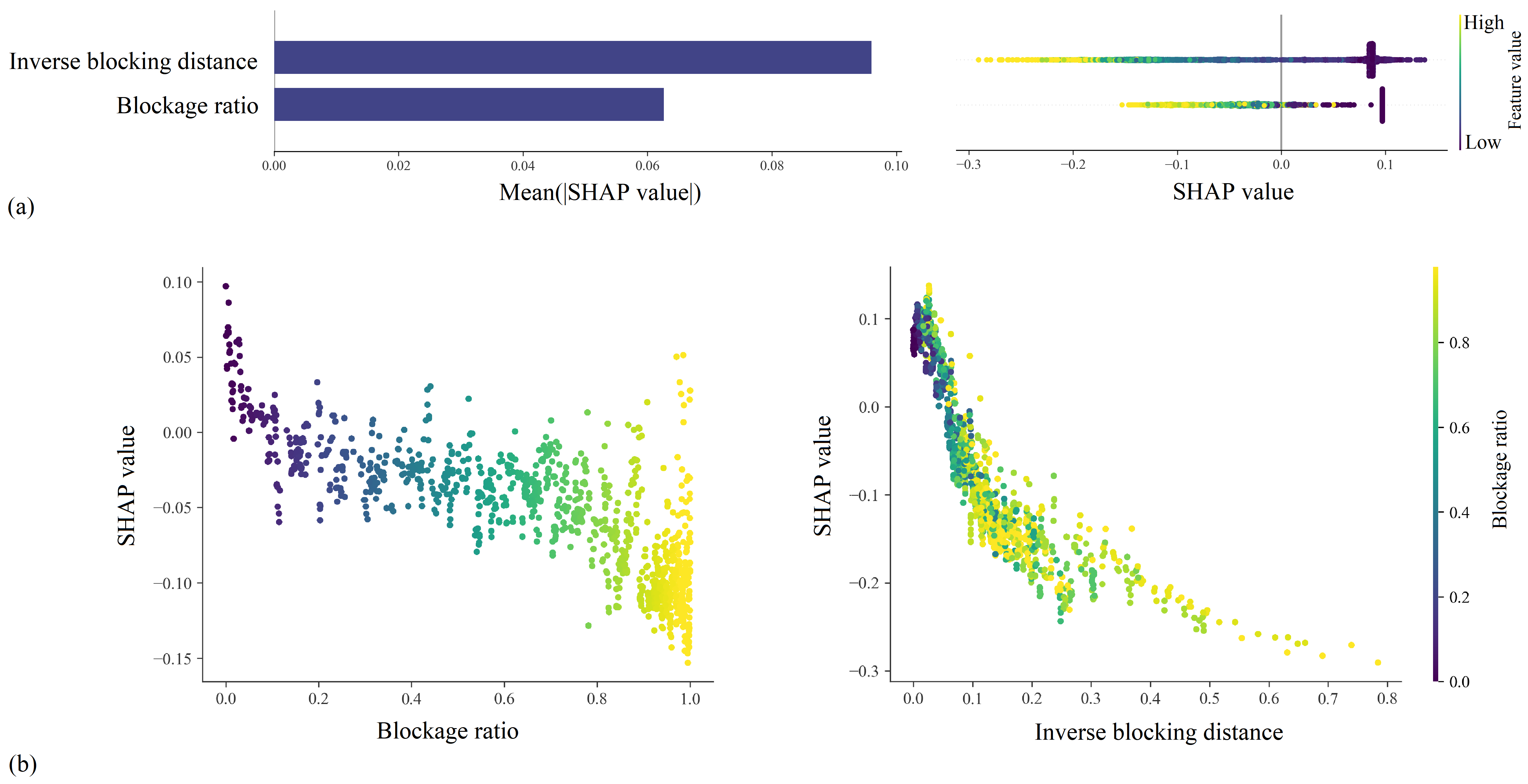}
	\caption{SHAP plots for the GM-ML model. (a) Left: Bar chart of the mean of the absolute SHAP value. Right: SHAP summary plot. Every instance of the dataset appears as a point. Each point is colored by the value of that feature and its position on the horizontal axis shows the impact of the feature on the model’s prediction for that example. When multiple points land at the same position, they stack vertically.  (b) SHAP dependence plot of the features.  }
	\label{fig:shapML1}
\end{figure*}

Figure~\ref{fig:shapML1} depicts the SHAP plots for the  GM-ML model. The absolute SHAP value for each feature across all the data (Figure~\ref{fig:shapML1}(a) on the left) shows that the inverse blocking distance is the most influential feature. To be able to understand how a feature’s value relates to its impact, and to identify the range and distribution of impacts that feature has on the model’s output, the SHAP summary plot ranked by mean absolute SHAP values (Figure~\ref{fig:shapML1}(a) on the right) is used. Focusing on the inverse blocking distance one can see that its impact on the model’s output smoothly varies as its value changes. In Figure~\ref{fig:shapML1}(b), the SHAP dependence plots of features are presented which give any feature versus its SHAP value. An inverse S-curve is observed for the blockage ratio revealing that the extreme values of this feature have a high impact on the predictions of the GM-ML model, while the intermediate values have a lower impact. 

\begin{figure*}[ht]
	\centering
	\includegraphics[width=1\textwidth]{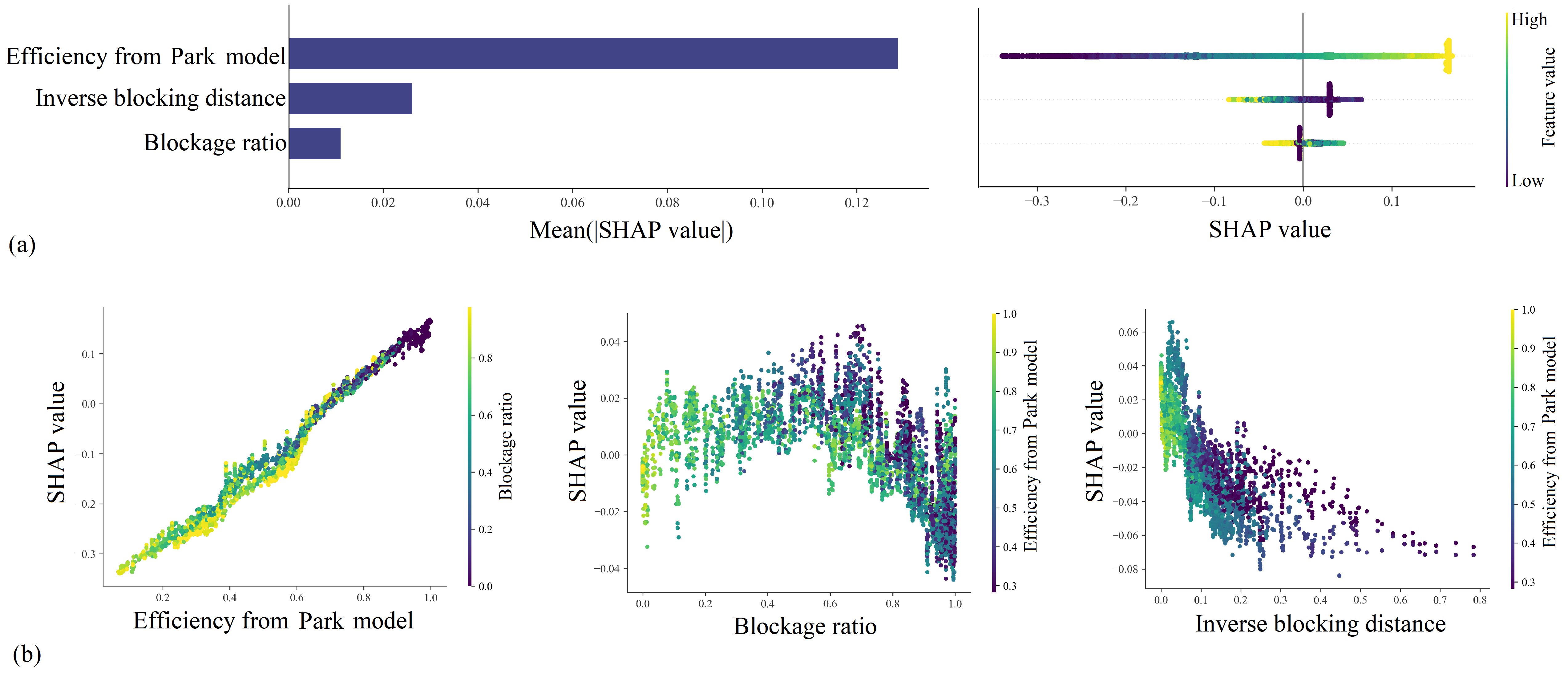}
	\caption{SHAP plots for Park-ML model. (a) Left: Bar chart of the mean of the absolute SHAP value. Right: SHAP summary plot. (b) Left: SHAP dependence plot of efficiency from the Park model. Center: SHAP dependence plot of blockage ratio. Right: SHAP dependence plot of inverse blocking distance. Points are colored based on their value of blockage ratio and the Park model's efficiency, respectively.}
	\label{fig:shapML2}
\end{figure*}

Figure~\ref{fig:shapML2} depicts the SHAP plots for the Park-ML model. As can be seen in Figure~\ref{fig:shapML2}(a), the turbine-level efficiency from the Park model is the most influential feature, and the blockage ratio is the least informative one. The impact of the most influential feature on the Park-ML model’s output smoothly varies as its value changes as shown in Figure~\ref{fig:shapML2}(a) (on the right). Figure~\ref{fig:shapML2}(b) which presents the  SHAP dependence plots of the features of this model shows that the impact of the blockage ratio on the model versus its value has a down-facing parabolic behavior, ascending behavior until the intermediate values of the blockage ratio followed by a descending behavior.

\begin{figure*}[ht]
	\centering
	\includegraphics[width=1\textwidth]{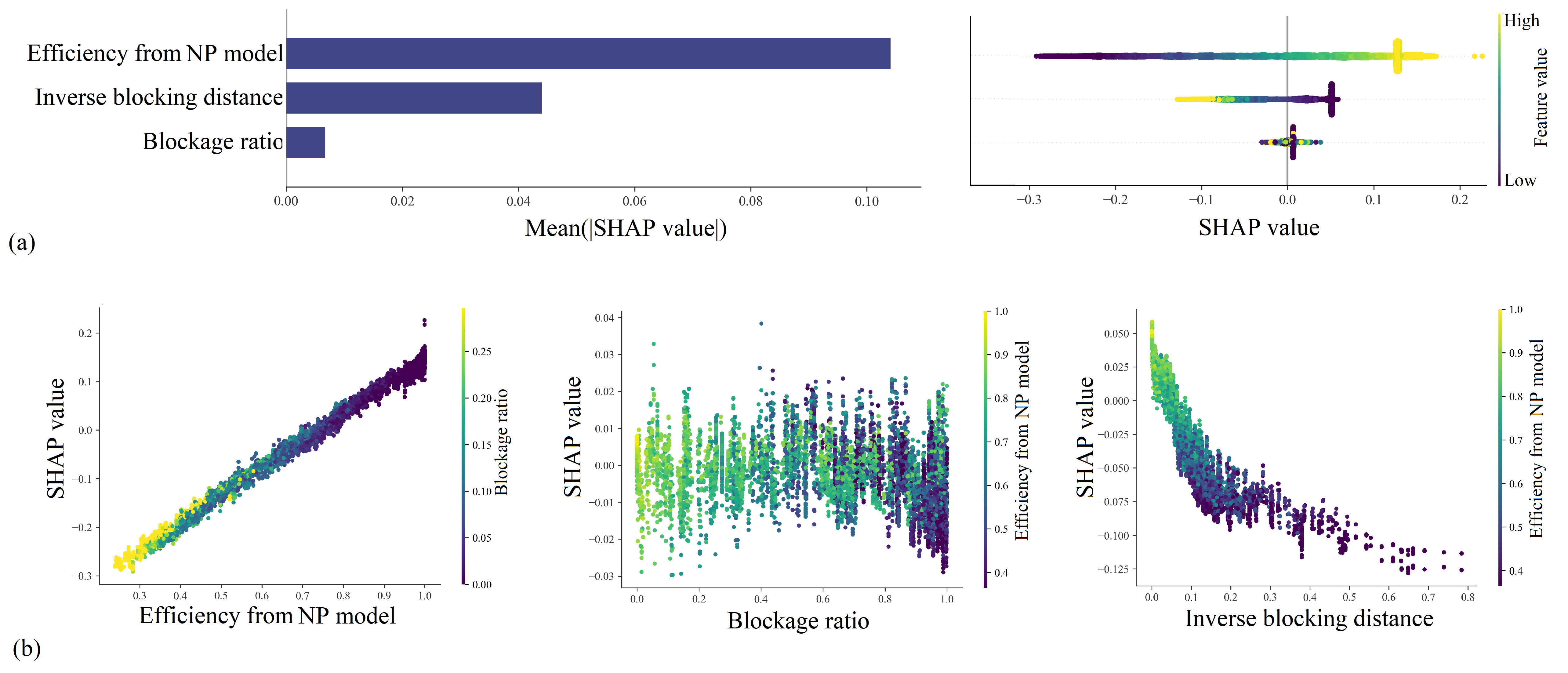}
	\caption{\blue SHAP plots for NP-ML model. (a) Left: Bar chart of the mean of the absolute SHAP value. Right: SHAP summary plot. (b) Left: SHAP dependence plot of efficiency from the NP model \cite{Niayifar2016}. Center: SHAP dependence plot of blockage ratio. Right: SHAP dependence plot of inverse blocking distance. Points are colored based on their value of blockage ratio and the NP model's efficiency, respectively.\black}
	\label{fig:shapML3}
\end{figure*}

\blue SHAP plots for the NP-ML model are presented in Figure~\ref{fig:shapML3}. By focusing on Figure~\ref{fig:shapML3}(a), one can see that similar to the Park-ML model, the efficiency from the physics-based model plays the most important role among the features. However, the inverse blocking distance has a larger impact on this model, and the points on the SHAP summary plot are vertically stacked at a higher SHAP value. The SHAP summary plot for the blockage ratio reveals that more points are centered around zero, and consequently, the mean absolute SHAP value of this feature is degraded compared to the Park-ML model. The SHAP dependence plots for the NP-ML model, given in Figure~\ref{fig:shapML3}(b), have similar behavior as those of the Park-ML model\black. 

\subsection{Machine-learning models' prediction on unseen cases}
Three cases of the generated data were left out so that they would not be seen by the ML models during the training phase. These three datasets are related to Cases (V), (VI), and (VII) that are different from those in training in terms of the turbines' operating conditions, inflow turbulence level, and the arrangement of the turbines to challenge the ML models from several points of view and to measure the success of guidance with physics in the generalization of the ML models. We compare our ML models with the Park model and the NP model \cite{Niayifar2016} by using the performance metric of turbine-level absolute percentage error (\textit{APE}) and its mean (\textit{MAPE}). Evaluation at the turbine level gives a better overview of models' performance since errors may fade when turbine efficiencies are aggregated to transform into farm efficiency. The \textit{APE} is defined as

\begin{align}
     APE &= \left\lvert \frac{{y_{\text{pred},j}-y_{\text{true},j}}}{y_{\text{true},j}} \right\rvert \times 100.
\end{align}

\begin{figure*}[ht!]
	\centering
	\includegraphics[width=1\textwidth]{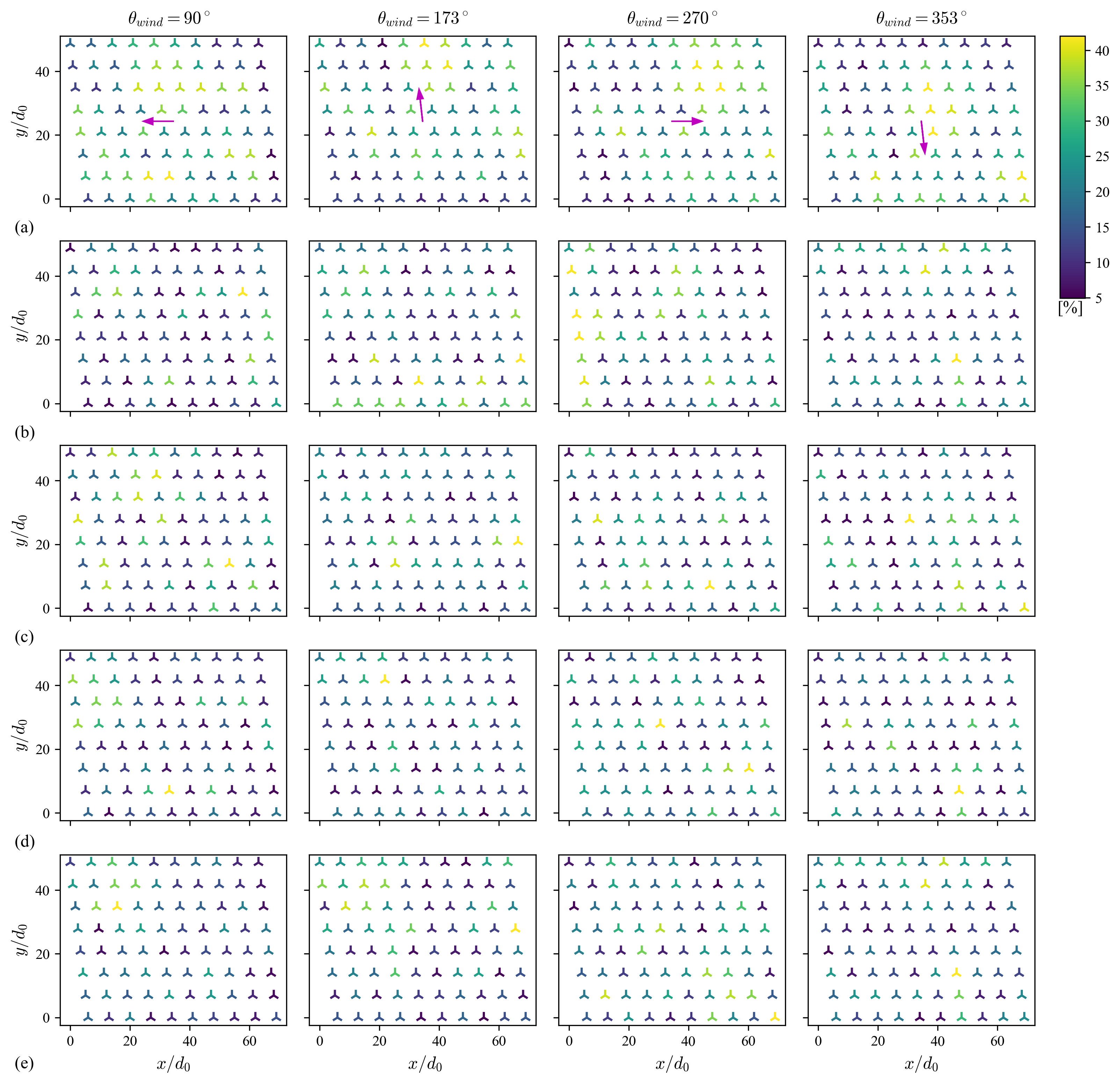}
	\caption{\blue The \textit{APE} of (a) Park model, (b) NP model \cite{Niayifar2016}, (c) GM-ML model, (d) Park-ML model, and (e) NP-ML model when applied to Case (V). The arrows indicate the wind direction corresponding to each column. \black}
	\label{fig:ml1HRrand}
\end{figure*}

Figure~\ref{fig:ml1HRrand} shows the \textit{APE} of Case (V) \blue for four different incoming wind directions\black, where the predictions of two analytical models, the Park model and the NP model \cite{Niayifar2016}, as well as the predictions of the GM-ML model, are compared with the physics-guided ML models, \blue i.e., the Park-ML and NP-ML models\black. The ML models outperform the Park model (Figure~\ref{fig:ml1HRrand}(a)) and the NP model \cite{Niayifar2016} (Figure~\ref{fig:ml1HRrand}(b)) for all four wind directions. The distribution of \textit{APE} for the predictions of the GM-ML model (Figure~\ref{fig:ml1HRrand}(c)) and the physics-guided ML models (Figure~\ref{fig:ml1HRrand}(d,e)) are somewhat similar. However, the number of turbines with higher \textit{APE} decreases when using the physics-guided ML models, \blue and since they have more complete physics of wake as their features, their predictions will be more reliable and more stable than that of the GM-ML model\black.

\begin{figure*}[ht!]
	\centering
	\includegraphics[width=0.53\textwidth]{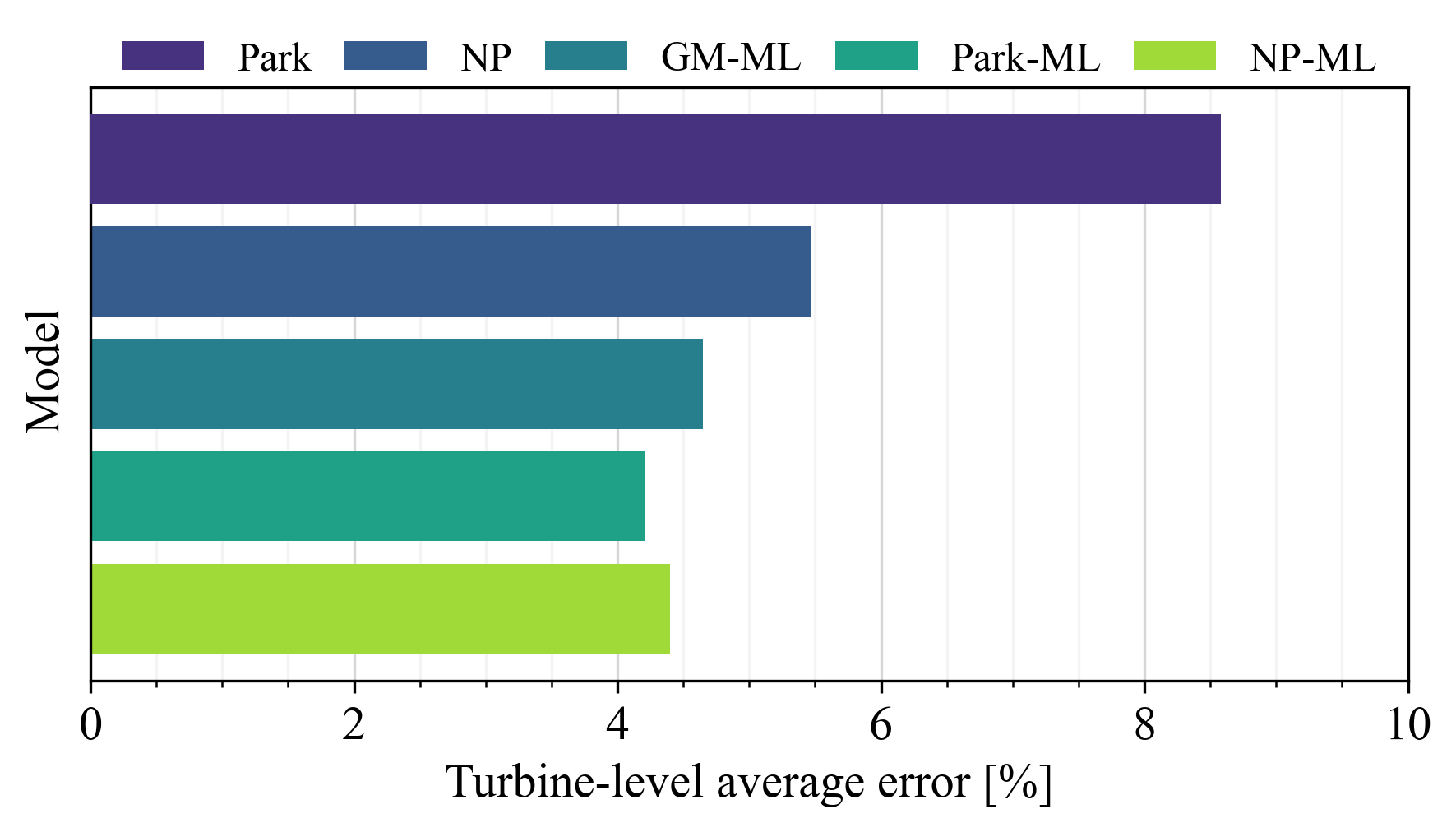}
	\caption{\blue The \textit{MAPE} of the physics-based wake models and the ML models averaged for all wind directions when applied to Case (V).\black}
	\label{fig:compBarRandHR}
\end{figure*}

For further evaluation of the performance of the different models in Case (V), their \textit{MAPE} indexes are compared in Figure~\ref{fig:compBarRandHR}. These errors are 8.58\% and 5.47\% for the Park model and NP model \cite{Niayifar2016}, respectively. The GM-ML model shows a \textit{MAPE} of 4.65\% when applied to Case (V) which outperforms both physics-based wake models. However, evaluation of the performance of the ML model with geometric features only in Cases (VI) and (VII) will give much more insight into it. 
\blue The Park-ML and NP-ML models are outperforming all other models with \textit{MAPE} of 4.21\% and 4.4\%, respectively. This interesting result shows that the physics-guided ML model is not sensitive to the selection of the physics-based model and is able to handle the complexity introduced due to the non-uniform distribution of thrust coefficient of turbines, while it was trained on cases with a uniform distribution of $C_T$. To investigate the dispersity of \textit{APE} in relation to the \textit{MAPE} of each model, we evaluate the standard deviation index that is equal to 7.11\%, 3.83\%, 3.6\%, 3.17\%, and 3.34\% for the Park-model, NP-model, GM-ML model, Park-ML model, and NP-ML model, respectively. \black

\begin{figure*}[ht!]
	\centering
	\includegraphics[width=1\textwidth]{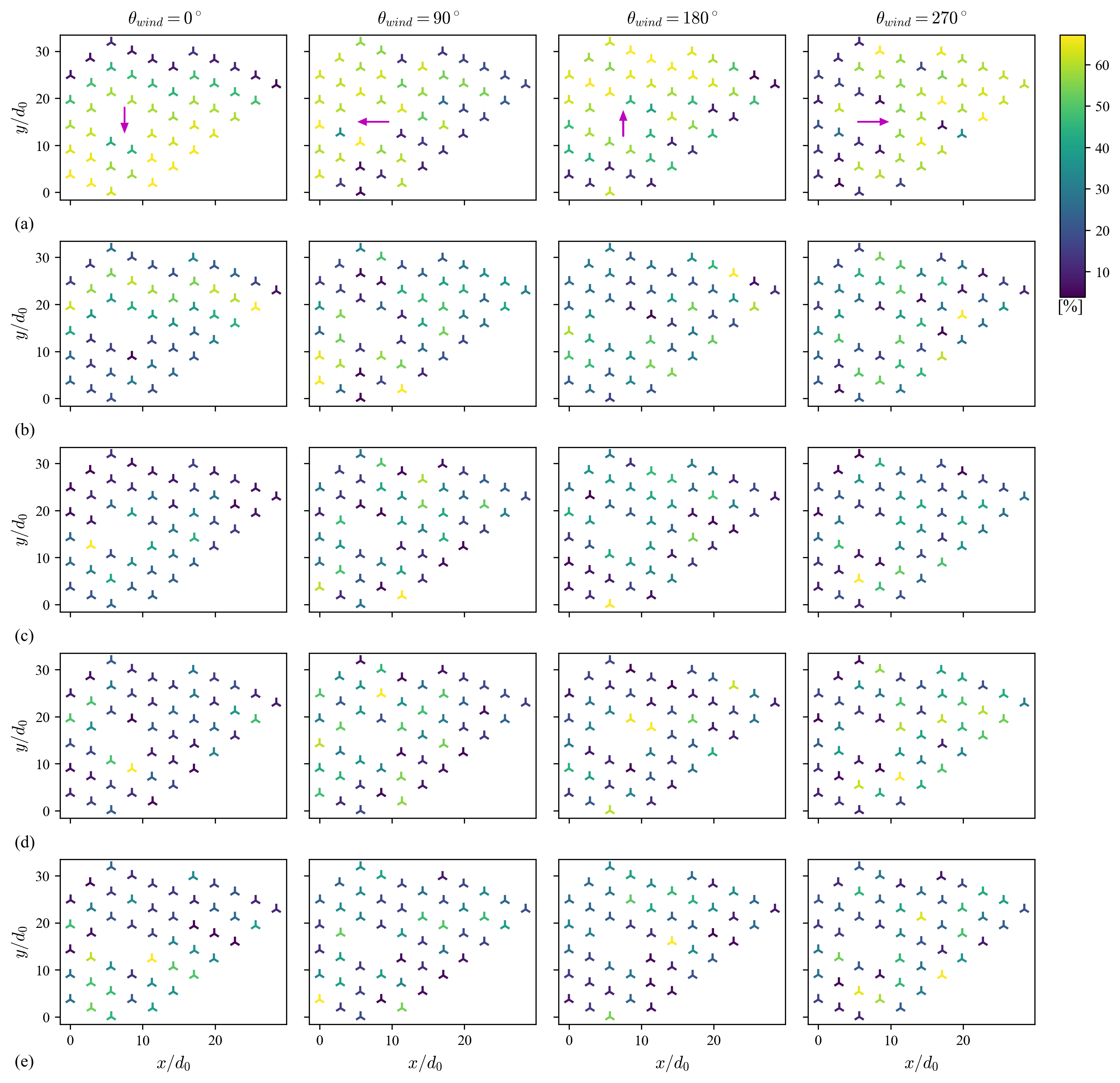}
	\caption{\blue The \textit{APE} of (a) Park model, (b) NP model \cite{Niayifar2016}, (c) GM-ML model, (d) Park-ML model, and (e) NP-ML model when applied to Case (VI). The arrows indicate the wind direction corresponding to each column. \black }
        \label{fig:lillgrund7p7MLturbine}
\end{figure*}

Case (VI), which is based on the layout of the Lillgrund offshore wind farm, has irregularities in the arrangement of turbines in its structure. On the other hand, the assumed thrust coefficient for its turbines is 0.86. This value of the thrust coefficient was not considered when preparing the data for the training stage, and if the model performs properly in estimating the performance of the turbines of this farm, it indicates the ability of the model in extrapolation. In Figure~\ref{fig:lillgrund7p7MLturbine}, the \textit{APE} of the turbines of this farm are analyzed for several wind directions. According to Figure~\ref{fig:lillgrund7p7MLturbine}(a), due to the high thrust coefficients of the turbines, the errors of the Park model are very high in predicting the efficiency of the downstream turbines, and this indicates an overestimation of the loss in the full wake by this model. 
Figure~\ref{fig:lillgrund7p7MLturbine}(b) is dedicated to errors in the estimation of the efficiency of the turbines using the NP model \cite{Niayifar2016}. This empirical-analytical model has correlations and assumptions based on experiments and LESs. It uses a linear superposition technique to consider wake interactions. Due to the higher degree of formulation complexity, it has a better performance compared to the Park model. 
By applying the GM-ML model to Case (VI), as shown in Figure~\ref{fig:lillgrund7p7MLturbine}(c), one can see that even though the performance is better than the Park model, a weaker performance is observed compared to the NP model \cite{Niayifar2016} \blue which originates from the fact the GM-ML model has no features corresponding to the operating conditions of turbines or inflow characteristics. Therefore, this results in the unstable and case-dependent behavior of the GM-ML model\black. 
Figure~\ref{fig:lillgrund7p7MLturbine}(d) shows the results of the  Park-ML model. What is clear is the significant reduction of the efficiency-prediction error compared to the Park model, which can indicate the effectiveness of the physics-guided ML model. The Park-ML model outperforms the  empirical-analytical NP model \cite{Niayifar2016} even though the operating conditions of turbines and the inflow turbulence level in Case (VI) are very close to the conditions that the parameters of the NP model \cite{Niayifar2016} were tuned for.
\blue Similarly, for the NP-ML model, the \textit{APE}, shown in Figure~\ref{fig:lillgrund7p7MLturbine}(e), are reduced compared to the corresponding physics-based model\black. 

\begin{figure*}[ht!]
	\centering
	\includegraphics[width=1\textwidth]{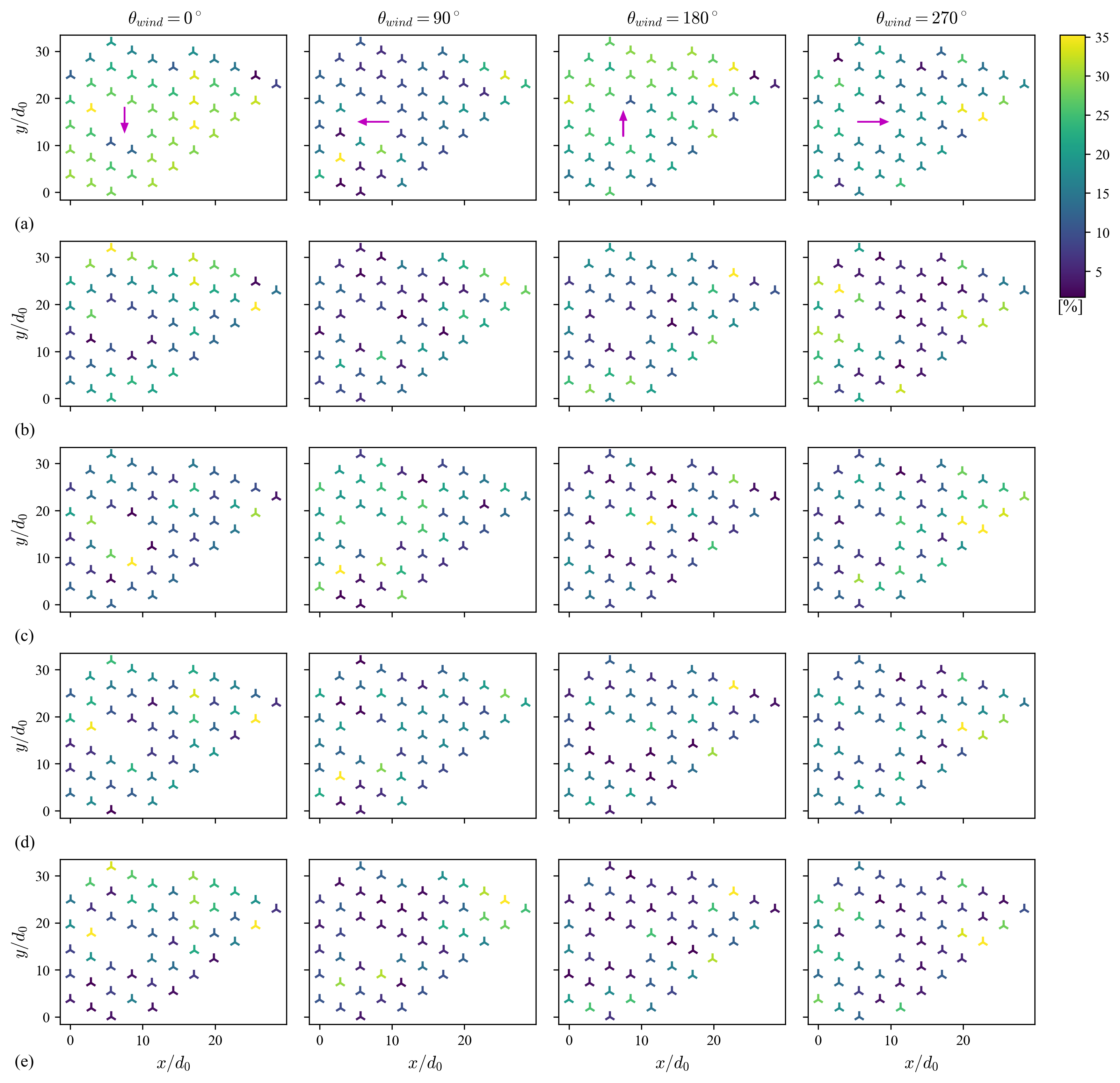}
	\caption{\blue The \textit{APE} of (a) Park model, (b) NP model \cite{Niayifar2016}, (c) GM-ML model, (d) Park-ML model, and (e) NP-ML model when applied to Case (VII). The arrows indicate the wind direction corresponding to each column. \black}
        \label{fig:lillgrund15p0MLturbine}
\end{figure*}

As described earlier, Case (VII) is different from the cases involved in the training phase in terms of turbine operating conditions, inflow turbulence level, and farm layout. The results of \textit{APE} for this case are shown in Figure~\ref{fig:lillgrund15p0MLturbine} revealing that the Park-ML \blue and NP-ML models are \black able to cope with such differences, extrapolate to a completely different case than what \blue they were \black trained upon, and outperform all the other models.  

\begin{figure*}[ht!]
	\centering
	\includegraphics[width=0.9\textwidth]{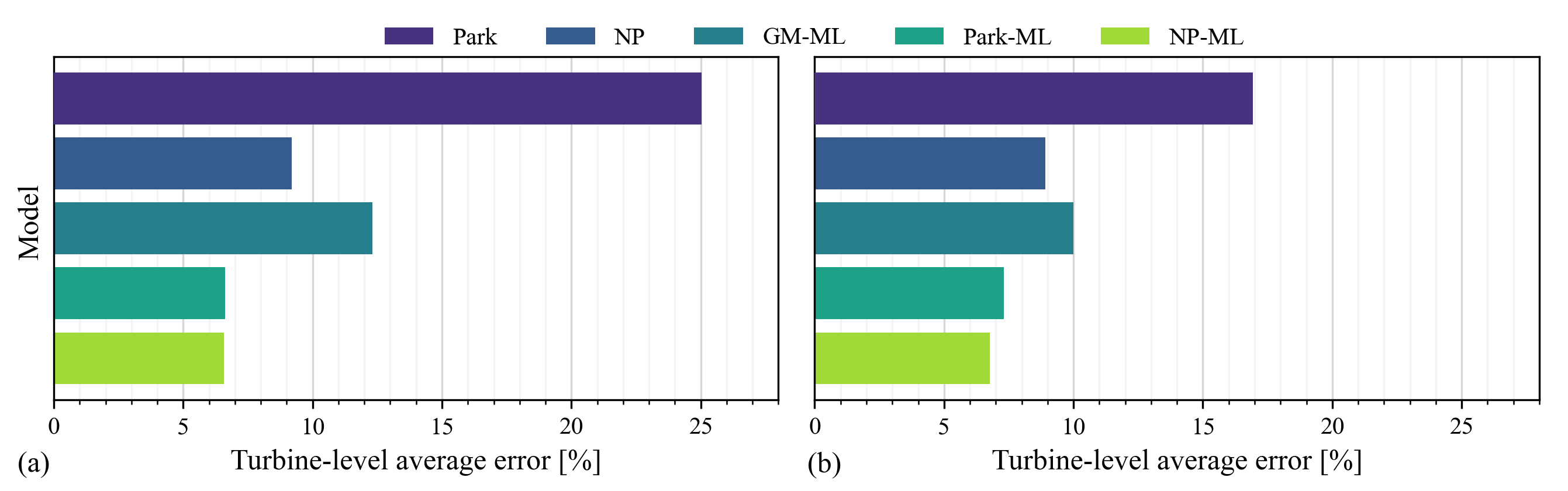}
	\caption{\blue The \textit{MAPE} of the physics-based models and the ML models averaged for all wind directions when applied to (a) Case (VI) and (b) Case (VII).\black}
	\label{fig:compBarLillgrund}
\end{figure*}

Figure~\ref{fig:compBarLillgrund} shows several interesting points that indicate the capability of the physics-guided ML models from two aspects of accuracy and generalization. The Park-ML \blue and NP-ML models have \black a lower error than the two physics-based models when applied to all unseen cases. In Case (VI), the \textit{MAPE} are 25.03\%, 9.19\%, 6.61\%, \blue and 6.58\% \black when using the Park model, the NP model \cite{Niayifar2016}, the Park-ML model, and the NP-ML model respectively. \blue The standard deviations of \textit{APE} of the mentioned models are  19.04\%, 6.30\%, 5.63\%, and 5.47\%, respectively\black. 
Applying the models to Case (VII) results in \textit{MAPE} equal to 16.94\%, 8.91\%, 7.3\%, \blue and 6.77\%, with standard deviations of 11.48\%, 6.47\%, 4.74\%, and 4.25\%, respectively. The interesting finding is that the physics-guided ML models show very minor sensitivity to the type of wake model used to guide them\black. Considering the different layout and operating conditions of Case (VI) from those in the training and prediction phase, and also the new inflow conditions of Case (VII), it can be concluded that the geometric features coupled with the efficiency values from the physics-based models have been able to equip the machine with generalization capabilities to a high extent. We also applied the GM-ML model to Cases (VI) and (VII), and the results show \textit{MAPE} of 12.32\% and 10.0\% \blue, respectively, with standard deviations equal to 11.24\% and 6.88\% for \textit{APE}\black. The GM-ML model outperforms the Park model in these two cases, but because of the physics-originated differences in Cases (VI) and (VII) compared to the cases in the training dataset, the GM-ML model cannot extrapolate to the same degree as the Park-ML model.

\section{Conclusions} \label{Sec:Conclusions}

The present study attempted to investigate the ability of turbine-level physics-guided ML models in wind-farm power prediction with the capacity to generalize to cases with unseen layouts, operating conditions, and inflow conditions as a proof of concept by focusing on the model itself and its features. Such models can initiate a similar path as we already have in physics-based model development, which starts from simple models and marches toward complex ones.  

To create the database required for the training and prediction phases, CFD simulations of seven cases based on the layouts of Horns rev 1 and Lillgrund offshore wind farms were performed for different incoming wind directions using a validated RANS framework. In the next step, \blue ML models were \black built using the XGBoost algorithm with three different physically interpretable features as their inputs to boost the machines' capability to extrapolate to newly introduced cases. The first set was comprised of two geometric features to characterize each turbine with respect to the upstream ones in different incoming wind directions, and the third feature was the estimated turbine efficiency using \blue two different physics-based models, i.e., the Park model and the empirical-analytical Gaussian wake model \cite{Niayifar2016}\black. The performance of the physics-guided ML models was compared against the Park model, the empirical-analytical Gaussian wake model \cite{Niayifar2016}, and another ML model with geometric features only, by applying them to three unseen cases. 
 
The results showed that the turbine-level physics-guided ML \blue models were \black able to estimate the performance of turbines with higher accuracy compared to the Park model, the empirical-analytical Gaussian wake model \cite{Niayifar2016}, and the ML model with geometric features only. \blue Another interesting finding was that the machine was not sensitive to the physics-based model chosen to guide it\black. The superior performance of the turbine-level physics-guided ML models when applied to unseen cases with different characteristics compared to the cases in the training phase showed their high level of generalizability. Guidance with physics and utilization of physically interpretable features helped the ML models to perform well and even satisfactorily extrapolate in cases with parameters outside the bounds that the model had seen in the training phase. The findings proved that this strategy can be utilized in developing accurate, lightweight, and generalizable data-driven models to predict the power of wind farms.

Future works can focus on the utilization of data from numerical simulations of various wind farms with a gradually increasing complexity, and available SCADA data of several operational wind farms to develop generalizable ML models following a progressive approach and building upon the ML models for unsteady wind-farm power prediction.

\section*{Acknowledgment}
The authors acknowledge the financial support from the Independent Research Fund Denmark (DFF) under Grant No. 0217-00038B.
\section*{Conflict of interest}
The authors have no conflicts to disclose.
\section*{Data availability statement}
The data that support the findings of this study are available from the corresponding author upon reasonable request.


\bibliography{Refs}

\end{document}